\documentclass[envcountsect,envcountsame,11pt]{llncs}
\usepackage[a4paper,hmargin=1in,vmargin=1.2in]{geometry}

\usepackage{latexsym}
\usepackage{amssymb}
\usepackage{amsmath}
\usepackage{float}
\usepackage{framed}
\usepackage{qip}
\usepackage{bbm}  
\usepackage{amsbsy}   
\usepackage{tabularx}
\usepackage{xspace}

\usepackage[active]{srcltx}

\newcommand{\RemoveAppendix}[1]{#1}     

\newcommand{\fv}[2]{#2} 

\newcounter{itm}
\newenvironment{myprotocol}[2][]
  {\begin{minipage}{\columnwidth} 
    \begin{framed}\hspace{0ex} 
     \begin{minipage}{0.99\columnwidth}
       {{\bf #2:} #1}
       \setcounter{itm}{1}
       \begin{list}{\arabic{itm}.}{\usecounter{itm}
          \setlength{\itemsep}{0mm}
          \setlength{\leftmargin}{\labelwidth}
          \setlength{\topsep}{\parsep}} \small }          
   {    \end{list}
       \vspace{-1.5ex} 
       \end{minipage} 
     \end{framed} 
    \end{minipage}\vspace{-0.6ex}}

\newenvironment{myfigure}[1]    
         {\begin{figure}[#1] \centering}
         { \end{figure}}

\allowdisplaybreaks

\renewcommand{\S}{\ensuremath{{\sf S}}} 
\renewcommand{\R}{\ensuremath{{\sf R}}} 
\newcommand{\dS}{\ensuremath{\tilde{\sf S}}} 
\newcommand{\dR}{\ensuremath{\tilde{\sf R}}} 

\newcommand{\V}{\ensuremath{{\sf V}}} 
\renewcommand{\C}{\ensuremath{{\sf C}}} 
\newcommand{\dC}{\ensuremath{\tilde{\sf C}}} 

\newcommand{\onetwo}[1][2]{\mbox{\textsl{1\hspace{-0.1ex}-#1}}}
\def\-{\hspace{-0.1ex}-\hspace{-0.1ex}}           
\newcommand{\pOT}{\textsl{OT}}                    
\newcommand{\OT}[1][2]{\textsl{\onetwo[#1]\:OT}}               

\newcommand{\Rand}{\textsl{Rand}}                 
\newcommand{\lStringOT}[1][2]{\textsl{\onetwo[#1]\:OT\,$^{\ell}$}\xspace}    
\newcommand{\RandlStringOT}[1][2]{\Rand\:\lStringOT[#1]}    

\newcommand{\Randqot}[1][2]{{\small \sc Rand 1\hspace{-0.1ex}-\hspace{-0.1ex}#1 QOT}\,}
\newcommand{\Randlqot}[1][2]{{\small \sc Rand 1\hspace{-0.1ex}-\hspace{-0.1ex}#1 QOT}\,$^{\ell}$}
\newcommand{\eprRandlqot}[1][2]{{\small\sc EPR Rand 1\hspace{-0.1ex}-\hspace{-0.1ex}#1 QOT}\,$^{\ell}$}

\newcommand{\ROT}{\textsl{ROT}}

\newcommand{\comm}{{\sc comm}}

\def\={\hspace{-0.5ex}=\hspace{-0.4ex}}          

\newcommand{\epsclose}{\approx_{\varepsilon}}

\newcommand{\eps}{\varepsilon}

\newcommand{\UH}{{\cal F}}      
\newcommand{\hf}{f}             
\newcommand{\Hf}{F}             

\newcommand{\set}[1]{\{#1\}}
\newcommand{\Set}[2]{\{ #1 : #2\}}

\newcommand{\univ}{two-universal\xspace}

\newcommand{\perm}[1]{\ensuremath{\Pi_n}}

\newcommand{\ol}[1]{\overline{#1}}

\newcommand{\assign}{\ensuremath{\kern.5ex\raisebox{.1ex}{\mbox{\rm:}}\kern -.3em =}}

\newcommand{\rs}{\regE}
\newcommand{\regE}{\ensuremath{\textit{\textsf{E}}}}

\newcommand{\nbit}{\{ 0,1 \} ^n}

\newcommand{\ball}[1]{{B}^{#1}}

\renewcommand{\I}{\mathbbm{1}}

\newcommand{\E}{\mathbb{E}}   

\renewcommand{\H}{\operatorname{H}} 
\newcommand{\Hb}{\H_{\mathrm{bin}}}   
\newcommand{\dist}{\operatorname{d}}   

\renewcommand{\mid}{\,|\,}

\spnewtheorem*{sketch}{Proof sketch}{\itshape}{\rmfamily}

\newcommand{\delete}[1]{}
\newcommand{\remove}[1]{}

\newcommand{\prei}[2][i-1]{#2^{#1}} 

\newcommand{\hmin}{\ensuremath{\H_{\infty}}}
\newcommand{\hmax}{\ensuremath{\H_{0}}}

\newcommand{\hie}[2]{\ensuremath{\hmin^{#1}(#2)}}

\newcommand{\hiee}[1]{\hie{\varepsilon}{#1}}
\newcommand{\ev}{\ensuremath{{\cal E}}\xspace}

\newcommand{\ep}{\varepsilon}

\newcommand{\dens}[1]{{\cal P}(#1)}

\newcommand*{\cB}{\mathcal{B}}
\newcommand*{\cH}{\mathcal{H}}
\newcommand*{\cU}{\mathcal{U}}
\newcommand*{\cX}{\mathcal{X}}
\newcommand*{\bbN}{\mathbb{N}}

\allowdisplaybreaks

\def\version$#1,v #2 #3/#4/#5 #6${#2 (#5-#4-#3)}

\title{A Tight High-Order Entropic Quantum Uncertainty Relation With
  Applications\thanks{This is the full version of~\cite{DFRSS07}}
 }

\newcounter{FICS}
\newcounter{SECOQC}

\newcommand{\ivan}{Ivan B. Damg{\aa}rd\inst{1}\fnmsep 
      \thanks{FICS,
      Foundations in Cryptography and Security,
      funded by the Danish Natural Sciences Research Council.}\setcounter{FICS}{\value{footnote}}
}

\newcommand{\serge}{Serge Fehr\inst{2}\fnmsep\thanks{Supported by 
the Dutch Organization for Scientific Research (NWO).}
}

\newcommand{\renato}{Renato Renner\inst{3}\fnmsep
 \thanks{Supported by HP Labs Bristol.}\fnmsep
 \thanks{Supported by the European project SECOQC.}\setcounter{SECOQC}{\value{footnote}}
 }

\newcommand{\louis}{Louis Salvail\inst{1} 
                              \fnmsep\thanks{QUSEP, Quantum Security in Practice, funded by the 
                                    Danish Natural Science Research Council.}
}

\newcommand{\christian}{Christian Schaffner\inst{2}\fnmsep
                               \footnotemark[\value{SECOQC}]
}

\institute{Basic Research in Computer
      Science (BRICS),
       funded by the Danish National Research Foundation,
       Department of Computer Science, University of Aarhus, Denmark, \email{\{ivan|salvail\}@brics.dk}.
       \and       
       Center for Mathematics and Computer Science (CWI),
     Amsterdam, Netherlands, \email{\{fehr|c.schaffner\}@cwi.nl}
      \and
     Cambridge University, UK,
     \email{r.renner@damtp.cam.ac.uk}
}

\author{\ivan \and \serge\ \and \renato \and \\ \louis \and \christian}

\titlerunning{High-Order Entropic Uncertainty Relation}
\authorrunning{I.~Damg\aa rd et al.}
\tocauthor{Ivan Damg\aa rd, Louis Salvail (University of AArhus),
           Serge Fehr, Christian Schaffner (CWI),
           Renato Renner (Cambridge)}

\begin{document}

\maketitle

\setcounter{footnote}{0}

\begin{abstract}
We derive a new entropic quantum uncertainty relation involving
min-entropy. The relation is tight and can be applied in various
quantum-cryptographic settings.

Protocols for quantum 1-out-of-2 Oblivious Transfer 
and quantum Bit Commitment 
are presented and the uncertainty relation is used to prove the
security of these protocols in the bounded-quantum-storage model
according to new strong security definitions.

As another application, we consider the realistic setting of Quantum
Key Distribution (QKD) against quantum-memory-bounded eavesdroppers.
The uncertainty relation allows to prove the security of QKD protocols
in this setting while tolerating considerably higher error rates
compared to the standard model with unbounded adversaries. For instance, for
the six-state protocol with one-way communication, a bit-flip error
rate of up to 17\% can be tolerated (compared to 13\% in the standard
model).

Our uncertainty relation also yields a lower bound on
the min-entropy key uncertainty against known-plaintext attacks
when quantum ciphers are composed. Previously, the 
key uncertainty  of these ciphers was only known with respect to
Shannon entropy. 
\end{abstract}



                                
\section{Introduction}

A 
problem often encountered in quantum cryptography is the
following: through some interaction between the players, a quantum
state $\rho$ is generated and then measured by
one of the players (call her Alice in the following). Assuming Alice is
honest, we want to know how 
unpredictable her measurement outcome is to the adversary.
Once a lower bound on the adversary's uncertainty about
Alice's measurement outcome is established, it
is usually easy to prove the desired security property of the
protocol. Many existing constructions in quantum cryptography have
been proved secure following this paradigm.

Typically,  Alice does not make her measurement in a fixed
basis, but chooses at random among a set of different bases. These
bases are usually chosen to be pairwise {\em mutually unbiased},
meaning that if $\rho$ is such that the measurement
outcome in one basis is fixed then this implies that the uncertainty
about the outcome of the measurement in the other basis is
maximal. In this way, one hopes to keep the
adversary's uncertainty high, even if $\rho$ is (partially)
under the adversary's control.

An inequality that lower
bounds the adversary's uncertainty in such a scenario
is called an {\em uncertainty relation}. 
There exist uncertainty relations for different measures of
uncertainty, but cryptographic applications typically 
require the adversary's min-entropy to be bounded from below.

In this paper, we introduce a new general and tight 
entropic uncertainty relation. Since the relation is expressed
in terms of high-order entropy (i.e. min-entropy), 
it is applicable to a large class of natural
protocols in quantum cryptography.
In particular, the new relation can
be applied in situations where an $n$-qubit state $\rho$ has each of
its qubits measured in a random and independent basis sampled
uniformly from a fixed set ${\cal B}$ of
bases. 
${\cal B}$ does not necessarily have to be mutually unbiased, but we
assume a 
lower bound $h$ (i.e. an {\em average entropic uncertainty bound}) on
the average Shannon entropy of the distribution $P_{\vartheta}$,
obtained by measuring an arbitrary $1$-qubit state in basis $\vartheta
\in {\cal B}$, meaning that $\frac{1}{|{\cal B}|}\sum_{\vartheta}
\H(P_{\vartheta}) \geq h$.

\medskip
\noindent
{\bf {\em Uncertainty Relation (informal):}} {\em Let $\cal B$ be a
  set of bases with an average entropic uncertainty bound $h$ as
  above.  Let $P_{\theta}$ denote the probability distribution defined
  by measuring an arbitrary $n$-qubit state $\rho$ in basis $\theta
  \in {\cal B}^n$. For a $\theta \in_R {\cal B}^n$ chosen uniformly at random,
  it holds except with negligible probability 
that 
\begin{equation}\label{main}
\hmin(P_\theta) \gtrsim n h \enspace .
\end{equation}
}


\vspace{-2ex}
Observe that (\ref{main}) cannot be improved significantly since 
the min-entropy of a distribution is  
at most equal to the Shannon entropy. Our uncertainty relation is therefore
asymptotically tight when the bound $h$ is tight. 

Any lower bound on the Shannon entropy associated to a set
of measurements ${\cal B}$ can be used in (\ref{main}).   
In the special case where the 
set of bases is ${\cal B}=\{+,\times\}$ (i.e. the two BB84 bases),
$h$ is known precisely using Maassen and Uffink's
entropic relation, see inequality~\eqref{maassenuffink} below.
We get $h=\frac{1}{2}$ and (\ref{main}) results in
$\hmin(P_{\theta}) \gtrsim \frac{n}{2}$.
Uncertainty relations for the BB84 coding scheme~\cite{BB84}
are useful since this coding is widely used in quantum cryptography.
Its resilience to imperfect
quantum channels, sources, and detectors is an important advantage
in practice.

We now discuss applications of our high-order uncertainty relation to
important scenarios in cryptography: two-party cryptography, quantum
key distribution and quantum encryption.

\paragraph{Application I: Two-Party Cryptography in the Bounded-Quantum-Storage Model.}
Entropic uncertainty relations are powerful
tools for the security analysis of cryptographic protocols  in 
the bounded-quantum-storage model. In this model, the adversary is unbounded
in every respect, except that at a certain time, his quantum memory is reduced
to a certain size (by performing some measurement). In~\cite{DFSS05}, 
an uncertainty relation involving min-entropy was shown and used
in the analysis of protocols for Rabin oblivious
transfer (\ROT)  and bit commitment. 
This uncertainty relation only applies in the case when $n$ qubits are all measured in one
out of two mutually unbiased bases.

A major difference between our result (\ref{main}) and the one
from~\cite{DFSS05} is that while both relations bound the min-entropy
conditioned on an event, this event happens in our case with
probability essentially 1 (on average) whereas the corresponding event
from~\cite{DFSS05} only happens with probability about 1/2.
In Sect.~\ref{sec:12OT}, we prove the following:

\medskip

\noindent
{\bf {\em \OT[2]\ in the Bounded-Quantum-Storage Model:}}
{\em
There exists a non-interactive protocol for 1-out-of-2 oblivious transfer (\OT[2])
of $\ell$-bit messages, secure 
against
adversaries with quantum memory size at most \mbox{$n/4 - 2\ell$}. Here, $n$ is the number of qubits
transmitted in the protocol and $\ell$ can be a constant fraction of $n$. Honest players need
no quantum memory.}
\medskip

Since all flavors of \pOT\ are known to be equivalent under classical
information-theoretic reductions, and a \ROT\ protocol is already
known from~\cite{DFSS05}, the above result may seem insignificant.
This is not the case, however, for several reasons: First, although it
may in principle be possible to obtain a protocol for \OT[2] from the \ROT\ 
protocol of~\cite{DFSS05} using the standard black-box reduction, the
fact that we need to call the \ROT\ primitive many times would force the bound on the adversary's memory to be {\em sub}linear (in the number of transmitted qubits). 
Second, the
techniques used in~\cite{DFSS05} do not seem applicable to \OT[2],
unless via the inefficient generic reduction to \ROT.  And, third, we
prove security according to a 
stronger definition than the one
used in~\cite{DFSS05}, namely a quantum version of a recent
classical definition for information theoretic \OT[2]~\cite{CSSW06}.
The definition ensures that all (dishonest) players' inputs are well
defined (and can be extracted when formalized appropriately).
In particular, this implies security under sequential composition
whereas composability of the protocol from~\cite{DFSS05} was not
proven.

Furthermore, our techniques for \OT[2] imply almost directly a
non-interactive bit commitment scheme (in the bounded-quantum-storage
model) satisfying a composable security definition.  As an immediate
consequence, we obtain secure {\em string} commitment schemes.  This
improves over the bit commitment construction of~\cite{DFSS05},
respectively its analysis, which does {\em not} guarantee
composability and thus does {\em not} necessarily allow for string
commitments. This application can be found in Sect.~\ref{sec:com}.



\paragraph{Application II: Quantum Key Distribution.}
We also apply our uncertainty relation to quantum key distribution
(QKD) settings.  QKD is the art of distributing a secret key between
two distant parties, Alice and Bob, using only a completely insecure
quantum channel and authentic classical communication. QKD protocols
typically provide information-theoretic security, i.e., even an
adversary with unlimited resources cannot get any information about
the key.  A major difficulty when implementing QKD schemes is
that they require a low-noise quantum channel.  The tolerated noise
level depends on the actual protocol and on the desired security of
the key. Because the quality of the channel typically decreases with
its length, the maximum tolerated noise level is an important
parameter limiting the maximum distance between Alice and Bob.

We consider a model in which the adversary has a limited amount of
quantum memory to store the information she intercepts during the
protocol execution. In this model, we show that the maximum tolerated
noise level is larger than in the standard scenario where the
adversary has unlimited resources.  For {\em one-way QKD protocols}
which are protocols where error-correction is performed
non-interactively (i.e., a single classical message is sent from one
party to the other), we show the following result:

\medskip
\noindent
{\bf {\em QKD Against Quantum-Memory-Bounded Eavesdroppers:}}
{\em
Let $\cB$ be a set of orthonormal bases of $\cH_2$ with average
entropic uncertainty bound $h$. Then, a \emph{one-way QKD-protocol} produces a secure key against
eavesdroppers whose quantum-memory size is sublinear in the length
of the raw key at a positive rate as long as the bit-flip probability
$p$ of the quantum channel fulfills $\Hb(p) < h $ where $\Hb(\cdot)$
denotes the binary Shannon-entropy function.
}
\medskip

Although this result does not allow us to improve (i.e. compared to
unbounded adversaries) the maximum error-rate for the BB84 protocol
(the four-state protocol), the six-state
protocol can be shown secure against adversaries with memory bound
sublinear in the secret-key length as long as the bit-flip error-rate
is less than $17\%$. This improves over the maximal error-rate of
$13\%$ for the same protocol against unbounded adversaries. We also
show that the generalization of the six-state protocols to more bases
(not necessarily mutually unbiased) can be shown secure (against
memory-bounded adversaries) for a maximal error-rate up to $20\%$
provided the number of bases is large enough. Note that the best
known one-way protocol based on qubits is proven secure against
general attacks for an error-rate of only up to roughly $14.1\%$, and
the theoretical maximum is $16.3\%$~\cite{RGK05}.
 
The quantum-memory-bounded eavesdropper model studied here is not
comparable to other restrictions on adversaries considered in the
literature (e.g. \emph{individual attacks}, where the eavesdropper is
assumed to apply independent measurements to each qubit sent over the
quantum channel~\cite{FGGNP97,lutkenhaus00}).  In fact, these
assumptions are generally artificial and their purpose is to simplify
security proofs rather than to relax the conditions on the quality of
the communication channel from which secure key can be generated.  We
believe that the quantum-memory-bounded eavesdropper model is more
realistic.

\paragraph{Application III: Key-Uncertainty of Quantum Ciphers.} 
In~\cite{DPS04}, symmetric quantum ciphers encrypting classical
messages with classical secret-keys are considered. It is shown that
under known-plaintext attacks, the Shannon uncertainty of the
secret-key can be much higher for some quantum ciphers than for any
classical one.  The Shannon secret-key uncertainty $\H(K|C,M)$ of
classical ciphers $C$ encrypting messages $M$ of size $m$ with keys
$K$ of size $k\geq m$ is always such that $\H(K|C,M)\leq k-m$.  In the
quantum case, the Shannon secret-key uncertainty is defined as the
minimum residual uncertainty about key $K$ given the best measurement
(POVM) $P_M(C)$ applied to quantum cipher $C$ given plaintext $M$.
Examples of quantum ciphers are provided with $k=m+1$ such that
$\H(K|P_M(C))=m/2+1$ and with $k=2m$ such that $\H(K|P_M(C))\geq
2m-1$.  All ciphers in~\cite{DPS04} have their keys consisting of two
parts.  The first part chooses one basis out a set ${\cal B}$ of bases
while the other part is used as a classical one-time-pad.  The message
is first encrypted with the one-time-pad before being rotated in the
basis indicated by the first part of the key. For one particular
cipher encrypting $m$-bit messages using $m+1$ bits
of key, Theorem~4 in~\cite{DPS04}
states that the Shannon secret-key uncertainty adds up under
repetitions with independent and random keys\footnote{ The proof of
  Theorem~4 in~\cite{DPS04} is incorrect but can easily be fixed\fv{}{
  without changing the statement}.}: if $\H(K|P_M(C))\geq h$ then $n$
repetitions with independent keys satisfy
$\H(K_1,\ldots,K_n|P_{M_1,\ldots,M_n}(C_1,\ldots,C_{n}))\geq n h$.
Our uncertainty relation allows to obtain a stronger result.  The
analysis in~\cite{DPS04} shows that these quantum ciphers with Shannon
secret-key uncertainty $h$ satisfy the condition of our uncertainty
relation. As result we obtain a lower bound on the min-entropy key
uncertainty given the outcome of any quantum measurement applied to
all ciphers and given all plaintexts.  When $\H(K|P_M(C))\geq h$ our
uncertainty relation tells us that
$\hmin(K_1,\ldots,K_n|P_{M_1,\ldots,M_n}(C_1,\ldots,C_n))\gtrsim nh$.
Notice that unlike the two previous applications, this time the result
holds without any restriction on the adversary. 

\paragraph{History and Related Work.}
The history of uncertainty relations starts with Heisenberg who showed
that the outcomes of two non-commuting observables $A$ and $B$ applied
to any state $\rho$ are not easy to predict simultaneously.  However,
Heisenberg only speaks about the variance of the measurement results.
Because his result had several shortcomings (as pointed out
in~\cite{HU88,Deutsch83}), more general forms of uncertainty relations
were proposed by Bialynicki-Birula and Mycielski~\cite{BM75} and by
Deutsch~\cite{Deutsch83}.  The new relations were called {\em entropic
  uncertainty relations}, because they are expressed using Shannon
entropy instead of the statistical variance and, hence, are purely
information theoretic statements. For instance, Deutsch's uncertainty
relation~\cite{Deutsch83} states that $\H(P)+\H(Q) \geq
-2\log{\frac{1+c}{2}}$, where $P,Q$ are random variables representing
the measurement results and $c$ is the maximum inner product norm
between any eigenvectors of $A$ and $B$.
First conjectured by Kraus~\cite{Kraus87}, Maassen and
Uffink~\cite{MU88} improved
Deutsch's relation to the optimal
\begin{equation}\label{maassenuffink}
\H(P)+\H(Q) \geq -2\log{c} \enspace .
\end{equation}

Although a bound on Shannon entropy can be helpful in some cases, it
is usually not good enough in cryptographic applications.  The main
tool to reduce the adversary's information|privacy
amplification~\cite{BBR88,ILL89,BBCM95,RK05,Renner05}|only works if a
bound on the adversary's min-entropy (in fact collision entropy) is
known.  Unfortunately, knowing the Shannon entropy of a distribution
does in general not allow to bound its higher order R\'enyi entropies.

An entropic uncertainty relation involving R\'enyi entropy of order
$2$ (i.e. {\em collision entropy}) was introduced by
Larsen~\cite{Larsen90,Ruiz95}.  Larsen's relation quantifies precisely
the collision entropy for the set $\{A_i\}_{i=1}^{d+1}$ of \emph{all}
maximally non-commuting observables, where $d$ is the dimension of the
Hilbert space.  Its use is therefore restricted to quantum coding
schemes that take advantage of \emph{all} $d+1$ observables, i.e. to
schemes that are difficult to implement in practice.  Uncertainty
relations in terms of R\'enyi entropy have also been studied in a
different context by Bialynicki-Birula~\cite{Bialynicki06}.


\section{Preliminaries}
\subsection{Notation and Terminology}\label{sec:Notation}
For any positive integer $d$, ${\cal H}_d$ stands for the complex
Hilbert space of dimension $d$ and $\dens{{\cal H}_d}$ for the set of
density operators, i.e., positive semi-definite trace-1 matrices,
acting on ${\cal H}_d$. The pair $\{\ket{0},\ket{1}\}$ denotes the
computational or rectilinear or ``$+$'' basis for the $2$-dimensional
Hilbert space ${\mathcal H}_2$.  The diagonal or ``$\times$'' basis is
defined as $\{\ket{0}_\times,\ket{1}_{\times}\}$ where
\smash{$\ket{0}_{\times}=(\ket{0}+\ket{1})/\sqrt2$} and
\smash{$\ket{1}_{\times}=(\ket{0}-\ket{1})/\sqrt2$}. The circular or
``$\oslash$'' basis consists of vectors \smash{$(\ket{0}+i
  \ket{1})/\sqrt2$} and \mbox{$(\ket{0} - i \ket{1})/\sqrt2$}.
Measuring a qubit in the $+\,$-basis (resp.\ $\times$-basis) means
applying the measurement described by projectors $\ket{0}\bra{0}$ and
$\ket{1}\bra{1}$ (resp.  projectors $\ket{0}_\times \bra{0}_\times$
and $\ket{1}_{\times}\bra{1}_\times$).  When the context requires it,
we write $\ket{0}_+$ and $\ket{1}_+$ instead of $\ket{0}$ and
$\ket{1}$, respectively.  If we want to choose the $+$ or
$\times$-basis according to the bit $b \in \{0,1\}$, we write $[ +,
\times ]_b$.


The behavior of a (mixed) quantum state in a register $\regE$ is fully
described by its density matrix~$\rho_\regE$. 
We often consider
cases where a quantum state may depend on some classical
random variable $X$, in that the state is described by the density
matrix $\rho_\regE^x$ if and only if $X = x$. For an observer who has
access to the state but not $X$, the behavior of the
state is determined by the density matrix $\rho_{\regE} :=\sum_x P_X(x)
\rho_\regE^x$, whereas the joint state, consisting of the classical $X$ and
the quantum register $\regE$ 
is described by the density matrix
$\rho_{X\regE} := \sum_x P_X(x) \proj{x} \otimes \rho_\regE^x$, where we understand
$\set{\ket{x}}_{x \in {\cal X}}$ to be the standard (orthonormal)
basis of ${\cal H}_{|{\cal X}|}$. Joint states with such \emph{c}lassical and 
\emph{q}uantum parts are called \emph{cq-states}.
We also write $\rho_X := \sum_x P_X(x) \proj{x}$ for the quantum
representation of the classical random variable $X$.
This notation extends naturally to quantum states that depend on
several classical random variables (i.e.~to ccq-states\fv{}{, cccq-states}
etc.).  Given a cq-state $\rho_{X \regE}$ as above, by saying that
there exists a random variable $Y$ such that $\rho_{XY\regE}$
satisfies some condition, we mean that $\rho_{X\regE}$ can be
understood as $\rho_{X\regE} = \tr_Y(\rho_{XY\regE})$ for some
ccq-state $\rho_{XY\regE}$ and that $\rho_{XY\regE}$ satisfies the
\fv{}{required} condition.\fv{}{\footnote{The quantum version is similar to the
  case of distributions of classical random variables where given $X$,
  the existence of a certain $Y$ is understood that there exists a
  joint distribution $P_{XY}$ with $\sum_y P_{XY}(\cdot,y) = P_X$. }}

We would like to point out that 
$\rho_{X\regE} = \rho_X \otimes \rho_\regE$ holds if and only if the
quantum part is independent of $X$ (in that $\rho_\regE^x =
\rho_\regE$ for any $x$), where the latter in particular implies that
no information on $X$ can be learned by observing only $\rho_{\regE}$.
Similarly, $X$ is uniformly random and independent of the quantum
state in register $\regE$ if and only if \smash{$\rho_{X\regE} =
  \frac{1}{|{\cal X}|}\I \otimes \rho_\regE$}, where
\smash{$\frac{1}{|{\cal X}|}\I$} is the density matrix of the fully
mixed state of suitable dimension.  Finally, if two states like
$\rho_{X\regE}$ and $\rho_X \otimes \rho_\regE$ are $\eps$-close in
terms of their trace distance $\delta(\rho,\sigma) = \frac{1}{2}
\tr(|\rho-\sigma|)$, which we write as $\rho_{X\regE} \epsclose \rho_X
\otimes \rho_\regE$, then the real system $\rho_{X\regE}$ ``behaves''
as the ideal system $\rho_X \otimes \rho_\regE$ except with
probability~$\varepsilon$ in that for any evolution of the system no
observer can distinguish the real from the ideal one with advantage
greater than~$\eps$~\cite{RK05}.

\subsection{Smooth R\'enyi Entropy}

We briefly recall the notion of (conditional) {\em smooth}
min-entropy~\cite{Renner05,RW05}. For more details, we refer to the
aforementioned literature.  Let $X$ be a random variable over alphabet
${\cal X}$ with distribution $P_X$. The standard notion of min-entropy is given by \mbox{$\hmin(X) = - \log \bigl(\max_x P_X(x)\bigr)$} and that of max-entropy by \mbox{$\hmax(X) = \log\big|\{x\in{\cal X}:P_X(x)>0\}\big|$}.
More general, for any event $\ev$ (defined by $P_{\ev|X}(x) = \Pr[\ev|X\!=\!x]$ for all
$x\in{\cal X}$) $\hmin(X\ev)$ may be defined similarly simply by
replacing $P_X$ by $P_{X\ev}$. Note that the ``distribution''
$P_{X\ev}$ is not normalized; $\hmin(X\ev)$ is still well defined,
though.  For an arbitrary $\varepsilon \geq 0$, the smooth version
$\hiee{X}$ is defined as follows.  $\hiee{X}$ is the {\em maximum} of
the standard min-entropy $\hmin(X\ev)$, where the maximum is taken
over all events $\ev$ with $\Pr(\ev) \geq 1 - \varepsilon$.
Informally, this can be understood that if $\hiee{X} = r$ then the
standard min-entropy of $X$ equals $r$ as well, except with
probability $\varepsilon$. As $\varepsilon$ can be interpreted as an
error probability, we typically require $\varepsilon$ to be negligible
in the security parameter $n$.

For random variables $X$ and $Y$, the {\em conditional}
smooth min-entropy $\hiee{X \mid Y}$ is defined as $\hiee{X \mid Y} =
\max_{\ev} \min_y \hmin(X\ev \mid Y\!=\!y)$, where the quantification
over $\ev$ is over all events $\ev$ (defined by $P_{\ev|X Y}$) with
$\Pr(\ev) \geq 1 - \varepsilon$.  In Sect.~\ref{sec:qkd}, we work with
smooth min-entropy conditioned on a quantum state. We refer the reader
to~\cite{Renner05} for the definition\fv{}{ of this quantum version}.  We
will make use of the following chain rule for smooth
min-entropy~\cite{RW05}, which in spirit was already 
shown in~\cite{Cachin97}.

\begin{lemma}\label{lem:chain}
$\hie{\varepsilon+\varepsilon'}{X \mid Y} > \hie{\varepsilon}{XY}
- \hmax(Y) - \log{\left(\frac{1}{\varepsilon'}\right)}$ for all 
$\varepsilon,\varepsilon' > 0$.
\end{lemma}



\subsection{Azuma's Inequality}\label{sec:Azuma}

In the following and throughout the paper, the expected value of a real-valued random variable $R$ is denoted by $\E[R]$. 
Similarly, $\E[R|{\cal E}]$ and $\E[R|S]$ denote the conditional expectation of $R$ conditioned on an event $\cal E$ respectively random variable~$S$.
\begin{definition}
A list of real-valued
random variables $R_1,\ldots,R_n$ is called
a \emph{martingale difference sequence} if $\E[R_i \mid R_1,\ldots, R_{i-1}] = 0$ with probability 1 for every $1 \leq i \leq n$, i.e., if
\mbox{$
\E[R_i \mid R_1\!=\!r_1,\ldots,R_{i-1}\!=\!r_{i-1}] =0
$}
for\fv{}{every} $1 \leq i \leq n$ and all 
$r_1,\ldots,r_{i-1} \in \mathbb{R}$.
\end{definition}
The following lemma follows directly from Azuma's
inequality~\cite{Azuma67,AS00}.
\begin{lemma}\label{lem:azuma}
Let $R_1, \ldots, R_n$ be a martingale difference sequence such that
$|R_i| \leq c$ for every $1 \leq i \leq n$. Then, 
$\Pr\bigl[ \sum_i R_i \geq \lambda n \bigr] \leq
\exp\bigl(-\frac{\lambda^2 n}{2 c^2}\bigr)$ for any $\lambda > 0$. 
\end{lemma}

\section{The Uncertainty Relation}
We start with a classical tool which itself might be of independent
interest.
\begin{theorem}\label{thm:hmin}
Let $Z_1,\ldots, Z_n$ be $n$ (not necessarily
independent) random variables over alphabet ${\cal Z}$, and let $h \geq 0$ be such that 
\begin{equation} \label{eq:entropyassumption}
\H(Z_i \mid Z_1=z_1,\ldots,Z_{i-1}=z_{i-1})\geq h
\end{equation}
 for all $1\leq i\leq n$ and $z_1,\ldots, z_{i-1} \in {\cal Z}$. 
Then for any $0<\lambda<\frac12$ 
\[
\hiee{Z_1,\ldots,Z_n} \geq (h-2\lambda) n \enspace ,
\]
where $\varepsilon =
\exp\bigl(-\frac{\lambda^2 n}{32\log(|\mathcal{Z}|/\lambda)^2}\bigr)$.
\end{theorem}
If the $Z_i$'s are {\em independent} and have Shannon-entropy at least $h$,
it is known (see~\cite{RW05}) that the smooth min-entropy of
$Z_1,\ldots,Z_n$ is, to good approximation, at least $n h$ for large enough~$n$.\fv{}{\footnote{An even weaker version is the so-called Flattening Lemma~\cite{GolVad99}, which requires the $Z_i$'s to be independent and {\em equally} distributed, with a given {\em lower bound} on the smallest probability. It is in particular this missing lower bound that makes our proof technically more involved. }} 
Informally,
Theorem~\ref{thm:hmin} guarantees that when the independence-condition
is relaxed to a lower bound on the Shannon entropy of $Z_i$ \emph{given
  any previous history}, then we still have min-entropy of (almost) $n h$ 
except with negligible probability~$\varepsilon$.
\begin{proof}[sketch]
  The idea is to use\fv{}{ Azuma's inequality in the form of}
  Lemma~\ref{lem:azuma} for cleverly chosen $R_i$'s.  For any $i$ we
  write $\prei[i]{Z} \assign (Z_1,\ldots,Z_i)$ (with $Z^0$ being the
  ``empty symbol"), and similarly for other sequences.  We want to
  show that \smash{$\Pr\!\big[P_{\prei[n]{Z}}(\prei[n]{Z}) \geq
    2^{-(h-2\lambda)n}\big] \leq \varepsilon$}.
By the definition of smooth min-entropy, this then implies the claim. 
Note that
$P_{\prei[n]{Z}}(\prei[n]{Z}) \geq 2^{-(h-2\lambda)n}$ is equivalent
to
\begin{equation*}
\sum_{i=1}^n \Big( \log\bigl( P_{Z_i \mid \prei{Z}}(Z_i \mid \prei{Z})
\bigr) + h \Big) \geq 2 \lambda n \enspace .
\end{equation*}

We set $S_i \assign \log P_{Z_i|\prei{Z}}(Z_i \mid \prei{Z})$.
For such a sequence of real-valued
random variables $S_1,\ldots,S_n$, it is easy to verify that
$R_1,\ldots,R_n$ where $R_i \assign S_i - \E[S_i \mid \prei{S}]$ forms a
martingale difference sequence. If the $|R_i|$ were bounded by $c$, we
could use Lemma~\ref{lem:azuma} to conclude that
\begin{equation*}
\Pr\Biggl[\sum_{i=1}^n \Big(S_i - \E\big[S_i \mid \prei{S}\big]\Big) \geq
  \lambda n\Biggr] \leq 
\exp\biggl(-\frac{\lambda^2 n}{2 c^2}\biggr) \enspace .
\end{equation*}
As by assumption $\E[S_i \mid \prei{S}] \leq -h$, this would give us a
bound similar to what we want to show. In order to enforce a bound on
$|R_i|$, $S_i$ needs to be truncated whenever $P_{Z_i \mid
  \prei{Z}}(Z_i \mid \prei{Z})$ is smaller than some $\delta > 0$.  It
is then a subtle and technically involved matter of choosing $\delta$
and $\varepsilon$ appropriately in order to finish the proof, as shown
in \fv{the full version of the
  paper~\cite{DFRSS07full}}{Appendix~\ref{app:proofhmin}}.\qed
\end{proof}

We now state and prove the new entropic uncertainty relation
in its most general form. A special case will then be introduced
(Corollary~\ref{cor:uncertainty})
and used in the security analysis of all protocols we consider
in the following. 
\begin{definition}\label{def:aeub}
  Let $\cal B$ be a finite set of orthonormal bases in the $d$-dimensional
  Hilbert space~${\cal H}_d$.  We call $h \geq 0$ an {\em average
    entropic uncertainty bound} for $\cal B$ if every state in ${\cal
    H}_d$ satisfies $\frac{1}{|{\cal B}|} \sum_{\vartheta \in \cal
    B}\H(P_{\vartheta}) \geq h$, where $P_{\vartheta}$ is the
  distribution obtained by measuring the state in basis $\vartheta$.
\end{definition}
Note that by the convexity of the Shannon entropy $\H$, a lower bound
for all \emph{pure} states in ${\cal H}_d$ suffices to imply the bound
for all (possibly mixed) states.

\begin{theorem}\label{thm:genrel}
  Let $\cal B$ be a set of orthonormal bases in ${\cal H}_d$ with an
  average entropic uncertainty bound $h$, and let 
  $\rho \in \dens{{\cal H}^{\otimes n}_d}$ be an arbitrary quantum state.
Let $\Theta = (\Theta_1,\ldots,\Theta_n)$ be uniformly distributed
over ${\cal B}^n$ and let $X = (X_1,\ldots,X_n)$ be the outcome when
measuring $\rho$ in basis $\Theta$, taking values from
\mbox{$\set{0,\ldots, d-1}^n$}.  Then for any $0 < \lambda < \frac12$,
$$
\hie{\varepsilon}{X \mid \Theta} \geq \left(h-
  2\lambda\right)n 
$$
with 
$\varepsilon = \exp \!\left( - \frac{\lambda^2 n}{32
    \left(\log(|{\cal B}|\cdot d / \lambda) \right)^2} \right)$.
\end{theorem}

\begin{proof}
For $i \in \set{1,\ldots,n}$ define $Z_i \assign (X_i,\Theta_i)$ and $\prei[i]{Z} \assign (Z_1,\ldots,Z_i)$. Let 
$\prei[i-1]{z}$ be arbitrary in $(\set{0,\ldots,d-1}\times{\cal B})^{i-1}$. Then
\begin{align*} 
\H(Z_i \mid \prei[i-1]{Z}\!=\!\prei[i-1]{z}) &= \H(X_i \mid \Theta_i, \prei[i-1]{Z}\!=\!\prei[i-1]{z})
+ \H(\Theta_i \mid \prei[i-1]{Z}\!=\!\prei[i-1]{z}) \geq h + \log{|{\cal B}|} \, ,
\end{align*}
where the inequality follows from the fact that $\Theta_i$ is chosen
uniformly at random and from the definition of $h$. Note that $h$
lower bounds the average entropy for any system in ${\cal H}_d$, 
and thus in particular for the $i$-th subsystem 
of $\rho$, with all
previous $d$-dimensional subsystems measured.  
Theorem~\ref{thm:hmin} thus implies that $\hie{\varepsilon}{X \Theta} \geq (h+\log|{\cal B}| - 2\lambda) n$ for any $0 < \lambda < \frac12$ and for $\eps$ as claimed. We conclude that 
\begin{align*}
\hie{\varepsilon}{X\mid \Theta}
&\geq \hie{\varepsilon}{X \Theta} - n\log|{\cal B}| \geq (h - 2 \lambda)n \enspace ,
\end{align*}
where the first inequality follows from the equality 
$$
P_{X\ev|\Theta}(x|\theta) = P_{X \Theta \ev}(x, \theta)/P_{\Theta}(\theta) = |{\cal B}|^n \cdot P_{X \Theta \ev}(x, \theta)
$$
for all $x$ and $\theta$ and any event $\ev$, and from the definition of (conditional) smooth entropy. \qed
\end{proof}

For the special case where ${\cal B}=\{+,\times\}$ is the set of BB84
bases, we can use the uncertainty relation of Maassen and
Uffink~\cite{MU88} (see (\ref{maassenuffink}) with $c=1/\sqrt{2}$),
which, using our terminology, states that
$\cal B$ has average entropic uncertainty bound $h =
\frac12$.  Theorem~\ref{thm:genrel} then immediately gives the
following corollary.

\begin{corollary}\label{cor:uncertainty}
  Let $\rho \in \dens{{\cal H}_2^{\otimes n}}$
  be an arbitrary $n$-qubit quantum state. Let $\Theta$ be uniformly distributed over
  $\set{+,\times}^n$, and let $X$ be the outcome
  when measuring $\rho$ in basis $\Theta$. Then for any $0 <
  \lambda < \frac12$,
$$
\hie{\varepsilon}{X \mid \Theta} \geq \left(\textstyle\frac{1}{2}
  - 2\lambda \right)n 
$$
where $\varepsilon = \exp\bigl( -\frac{\lambda^2 n}{32 \left(2 -
      \log(\lambda) \right)^2}\bigr) $.
\end{corollary}

Maassen and Uffink's relation being optimal means there exists a
quantum state $\rho$|namely the product state of eigenstates of the
subsystems, e.g.~$\rho=\proj{0}^{\otimes n}$|for which $\H(X \mid
\Theta) = \frac{n}{2}$. On the other hand, we have shown that
$(\frac12 - \lambda)n \leq \hiee{X \mid \Theta}$ for $\lambda>0$
arbitrarily close to $0$. For the product state $\rho$, the $X_i$'s
are independent and we know from~\cite{RW05} that in this case
$\hiee{X \mid \Theta}$ approaches $\H(X \mid \Theta) = \frac{n}{2}$.
%
It follows that the relation cannot be 
significantly improved even when considering R\'enyi entropy of lower order than min-entropy
(but higher than Shannon entropy). 

Another tight corollary is obtained if we consider the set of
measurements ${\cal B}=\{+,\times,\oslash\}$. In~\cite{Ruiz93},
S\'anchez-Ruiz has shown that for this ${\cal B}$ the average entropic
uncertainty bound $h=\frac{2}{3}$ is optimal. It implies that
$\hie{\varepsilon}{X\mid \Theta} \approx \H(X\mid \Theta) =
\frac{2n}{3}$ for negligible $\varepsilon$. In \fv{the full
  version~\cite{DFRSS07full}}{Appendix~\ref{app:uncertbound}}, we
compute the average uncertainty bound for the set of \emph{all bases}
of a $d$-dimensional Hilbert space.

\section{Application: Oblivious Transfer}\label{sec:12OT}

\subsection{Privacy Amplification and a Min-Entropy-Splitting Lemma}

Recall, a class $\UH$ of hash functions from, say, $\set{0,1}^n$ to
$\set{0,1}^{\ell}$ is called {\em \univ} \cite{WC77,WC79} if $\Pr[F(x)\!=\!F(x')] \leq 1/2^{\ell}$ for any
distinct $x,x' \in \set{0,1}^n$ and for $F$ uniformly distributed over
$\UH$.

\begin{theorem}[Privacy Amplification~\cite{RK05,Renner05}]\label{thm:pasmooth}
  Let $\varepsilon \geq 0$. Let $\rho_{XU \regE}$ be a ccq-state,
  where $X$ takes values in $\nbit$, $U$ in the finite domain $\cU$
  and register $\regE$ contains q qubits. Let $F$ be the random and
  independent choice of a member of a \univ\ class of hash functions
  $\UH$ from $\set{0,1}^n$ into $\set{0,1}^{\ell}$.
  Then, 
\begin{align}
  \delta\big(\rho_{F(X) F U \regE}, {\textstyle\frac{1}{2^{\ell}}\I }
  \otimes \rho_{F U \regE}\big) \leq \frac{1}{2} \,
  2^{-\frac{1}{2}\big(H_{\infty}^{\eps} (X|U)-q-\ell\big)} + 2 \eps \enspace . \label{dbound}
\end{align}
\end{theorem}
The theorem stated here is slightly different from the version given
in~\cite{RK05,Renner05} in that the classical and the quantum parts of
the adversary's knowledge are treated differently. A derivation of the
above theorem starting from the result in~\cite{Renner05} \fv{can be
  found in the full version~\cite{DFRSS07full}}{is given in
  Appendix~\ref{app:derivePA}}.

A second tool we need is the following Min-Entropy-Splitting Lemma. Note
that if the joint entropy of two random variables $X_0$ and $X_1$ is
large, then one is tempted to conclude that at least one of $X_0$ and
$X_1$ must still have large entropy, e.g.\ half of the original
entropy. Whereas this is indeed true for Shannon entropy, it is in
general not true for min-entropy. The following lemma, though, which
appeared in a preliminary version of~\cite{Wullschleger07}, shows that
it {\em is} true in a randomized sense. For completeness, the proof
can be found \fv{in the full version~\cite{DFRSS07full}}{in
  Appendix~\ref{app:ESL}}.

\begin{lemma}[Min-Entropy-Splitting Lemma]\label{lemma:ESL}
  Let $\varepsilon \geq 0$, and let $X_0,X_1$ be random variables (over possibly different alphabets) with
  \mbox{$\hmin^{\varepsilon}(X_0 X_1) \geq \alpha$}. 
Then, there exists 
a
  binary random variable $C$ over $\set{0,1}$ such that
  \mbox{$\hmin^{\varepsilon}(X_{1-C} C) \geq \alpha/2$}.
\end{lemma}

The corollary below follows rather straightforwardly by noting that
(for normalized as well as non-normalized distributions) $\hmin(X_0
X_1\mid Z) \geq \alpha$ holds exactly if $\hmin(X_0 X_1\mid Z\!=\!z)
\geq \alpha$ for all $z$, applying the Min-Entropy-Splitting Lemma,
and then using the Chain Rule, Lemma~\ref{lem:chain}.

\begin{corollary}\label{cor:ESL}
  Let $\ep \geq 0$, and let $X_0$, $X_1$ and $Z$ be random variables (over possibly different alphabets) such that
  $\hiee{X_0 X_1 \mid Z} \geq \alpha$. Then, there exists a binary
  random variable $C$ over $\set{0,1}$ such that
$\hie{\ep+\ep'}{X_{1-C} \mid Z C} \geq \alpha/2 - 1 - \log(1/\ep') \enspace$  for any $\ep' > 0$. 
\end{corollary}

\subsection{The Definition}\label{sec:rabin-obliv-transf}
In \lStringOT, the sender Alice sends two $\ell$-bit strings $S_0,
S_1$ to the receiver Bob in such a way that Bob can choose which
string to receive, but does not learn anything about the
other. On the other hand, Alice does not get to know which string Bob has chosen. The
common way to build \lStringOT is by constructing a protocol for
\mbox{(Sender-)}Randomized \lStringOT, which then can easily be
converted into an ordinary \lStringOT (see, e.g., \cite{DFSS06}).
\RandlStringOT\ essentially coincides with ordinary \lStringOT, except
that the two strings $S_0$ and $S_1$ are not \emph{input} by the
sender but generated uniformly at random during the protocol and
\emph{output} to the sender.



For the formal definition of the security requirements of a quantum protocol for
\RandlStringOT, let us fix the following notation: Let $C$
denote the binary random variable describing receiver \R's choice bit,
let $S_0, S_1$ denote the $\ell$-bit long random variables describing
sender \S's output strings, and let $Y$ denote the $\ell$-bit long
random variable describing \R's output string (supposed to be $S_C$).
 Furthermore, for a fixed candidate protocol for \RandlStringOT, and for a fixed
input distribution for $C$, the overall quantum state in case of a
dishonest sender $\dS$ is given by the ccq-state $\rho_{C Y \dS}$.
Analogously, in the case of a dishonest receiver \smash{$\dR$}, we have the
ccq-state $\rho_{S_0 S_1 \dR}$.



\begin{definition}[\RandlStringOT] \label{def:Rl12OT}
  An $\varepsilon$-secure {\em \RandlStringOT} is a quantum protocol
  between $\S$ and $\R$, with $\R$ having input $C \in \{0,1\}$ while
  $\S$ has no input, such that for any distribution of $C$, if\ \S\ and\ \R\ follow the
  protocol, then $\S$ gets output $S_0,S_1 \in \{0,1\}^{\ell}$ and
  $\R$ gets $Y = S_C$, except with probability~$\varepsilon$, and the
  following two properties hold:
\begin{description}
\item[\boldmath$\varepsilon$-Receiver-security:] If $\R$ is honest, then
  for any $\dS$, there exist 
  random variables $S'_0,S'_1$ such that \smash{$\Pr\bigl[Y=S'_C\bigr] \geq 1-\varepsilon$} and $\delta\bigl( \rho_{C S'_0 S'_1 \dS} , \rho_C \otimes \rho_{S'_0 S'_1 \dS}
  \bigr) \leq \eps$.
\item[\boldmath$\varepsilon$-Sender-security:] If $\S$ is honest, then for
  any $\dR$, there exists a binary random variable $C'$ such
  that $\delta\bigl( \rho_{S_{1-C'} S_{C'} C' \dR} , \frac{1}{|2^\ell|}\I \otimes \rho_{S_{C'}
      C' \dR} \bigr) \leq \eps$.
\end{description}
If any of the above holds for $\varepsilon = 0$, 
then the corresponding property is said to hold {\em perfectly}. 
If one of the properties only holds with respect to a restricted class
$\mathfrak{S}$ of \dS's respectively $\mathfrak{R}$ of \dR's, then this property
is said to hold and the protocol is said to be secure {\em against}
$\mathfrak{S}$ respectively~$\mathfrak{R}$.
\end{definition}

Receiver-security, as defined here, implies that whatever a dishonest
sender does is as good as the following: generate the ccq-state
\smash{$\rho_{S_0' S'_1 \dS}$} independently of $C$, let $\R$ know $S'_C$, and
output $\rho_{\dS}$.  On the other hand, sender-security implies that
whatever a dishonest receiver does is as good as the following:
generate the ccq-state \smash{$\rho_{S_{C'} C' \dR}$}, let $\S$
know $S_{C'}$ and an independent uniformly distributed $S_{1-C'}$, and
output $\rho_{\dR}$.  In other words, a protocol satisfying
Definition~\ref{def:Rl12OT} is a secure implementation of the natural
\RandlStringOT ideal functionality, except that it allows a dishonest
sender to influence the distribution of $S_0$ and $S_1$, and the
dishonest receiver to influence the distribution of the string of his
choice. This is in particular good enough for constructing a standard
\lStringOT in the straightforward way.

We would like to point out the importance of requiring the existence
of $S_0'$ and $S_1'$ in the formulation of receiver-security in a
quantum setting: requiring only that the sender learns no information
on $C$, as is sufficient in the classical setting
(see~e.g.~\cite{CSSW06}), does not prevent a dishonest sender from
obtaining $S_0,S_1$ by a suitable measurement {\em after} the
execution of the protocol in such a way that he can choose $S_0 \oplus
S_1$ at will, and $S_C$ is the string the receiver has obtained in the
protocol. 
\subsection{The Protocol}\label{sec:otprot}
We introduce a quantum protocol for \RandlStringOT that will be shown
perfectly receiver-secure against any sender and
$\varepsilon$-sender-secure against any quantum-memory-bounded
receiver for a negligible $\varepsilon$. The first two steps of the
protocol are identical to Wiesner's ``conjugate coding''
protocol~\cite{Wiesner83} from circa 1970 for \emph{``transmitting two
  messages either but not both of which may be received''}.

The simple protocol is described in Fig.~\ref{fig:Randlqot}, where for
 $x \in \nbit$ and  $I \subseteq \{1, \ldots,n\}$ 
 we define $x|_I$ to be the restriction of $x$ to the bits $x_i$ with $i \in I$. 
The sender $\S$ sends random BB84 states to the receiver $\R$, who measures all
received qubits according to his choice bit $C$. $\S$ then picks
randomly two 
functions from a fixed \univ class of hash functions $\UH$ from $\set{0,1}^n$ to $\set{0,1}^{\ell}$, where $\ell$ is to be
determined later, and applies them to the bits encoded in the $+$
respectively the bits encoded in $\times$-basis to obtain the output
strings $S_0$ and $S_1$. 
Note that we may apply a function $\hf \in \UH$ to a $n'$-bit string with $n' < n$ by padding it with zeros (which does not decrease its entropy). 
$\S$ announces the encoding bases and the hash
functions to the receiver who then can compute $S_C$. Intuitively, a
dishonest receiver who cannot store all the qubits until the right
bases are announced, will measure some qubits in the wrong basis and thus cannot learn both strings simultaneously.  

\begin{myfigure}{h}
\begin{myprotocol}[\small Let $c$ be $\R$'s choice bit.]{\Randlqot}
\item $\S$ picks $x \in_R \nbit$ and $\theta \in_R \{+,\times \}^n$, 
and sends $\ket{x_1}_{\theta_1},
   \fv{}{\ket{x_2}_{\theta_2}, }\ldots, \ket{x_n}_{\theta_n}$ to $\R$. 
\item $\R$ measures all qubits in basis $[+,\times]_{c}$. 
Let $x' \in \{0,1\}^n$ be the result.
\item $\S$ picks two hash functions $\hf_0,\hf_1 \in_R \UH$, announces $\theta$ and $\hf_0, \hf_1$ to $\R$, and
  outputs $s_0 \assign \hf_0(x |_{I_0})$ and $s_1 \assign
  \hf_1(x |_{I_1})$ where $I_b \assign \Set{i}{\theta_i \!=\! [+,\times]_b}$. \label{bound}
\item $\R$ outputs $s_{c} = \hf_{c}(x' |_{I_{c}})$.
\end{myprotocol}
\caption{Quantum Protocol for \boldmath\RandlStringOT.}\label{fig:Randlqot}
\end{myfigure}

We would like to stress that although protocol description and
analysis are designed for an ideal setting with perfect noiseless
quantum communication and with perfect sources and detectors, all our
results can easily be extended to a more realistic noisy setting along
the same lines as in~\cite{DFSS05}.

It is clear by the non-interactivity of \Randlqot\ that a
dishonest sender cannot learn anything about the receiver's choice
bit. Below, we show \Randlqot\ perfectly receiver-secure
according to Definition~\ref{def:Rl12OT}; the idea, though, simply is to have a dishonest $\dS$ execute the protocol with a receiver that has {\em unbounded quantum memory} and that way can compute $S'_0$ and $S'_1$. 
\begin{proposition}\label{prop:sec:oblivious}
\Randlqot\ is perfectly receiver-secure.
\end{proposition}

\begin{proof}
  Recall, the ccq-state $\rho_{C Y \dS}$ is defined by the experiment
  where $\dS$ interacts with the honest memory-bounded $\R$. We now
  define (in a new Hilbert space) the ccccq-state \smash{$\hat{\rho}_{\hat{C}
    \hat{Y} \hat{S}'_0 \hat{S}'_1 \dS}$} by a slightly different
  experiment: We let $\dS$ interact with a receiver with {\em
    unbounded} quantum memory, which waits to receive $\theta$ and
  then measures the $i$-th qubit in basis $\theta_i$ for
  $i=1,\ldots,n$. Let $X$ be the resulting string, and define
  $\hat{S}'_0 = f_0(X |_{I_0})$ and $\hat{S}'_1 = f_1(X|_{I_1})$.  Finally, sample $\hat{C}$ according to $P_C$ and set
  $\hat{Y} = \hat{S}'_C$.  It follows by construction that
  \smash{$\Pr\!\big[\hat{Y} \!\neq\! \hat{S}'_{\hat{C}}\big] = 0$} and
  $\hat{\rho}_{\hat{C}}$ is independent of $\hat{\rho}_{\hat{S}'_0
      \hat{S}'_1 \dS}$.  It remains to argue that $\hat{\rho}_{\hat{C}
      \hat{Y} \dS} = \rho_{C Y \dS}$, so that corresponding $S'_0$ and
    $S'_1$ also exist in the original experiment. But this is
    obviously satisfied since the only difference between the two
    experiments is when and in what basis the qubits at position $i
    \in I_{1-C}$ are measured, which does not affect $\rho_{C Y \dS}$ respectively $\hat{\rho}_{\hat{C} \hat{Y} \dS}$. 
\end{proof}



We model dishonest receivers in \Randlqot\ under the assumption that
the maximum size of their quantum storage is bounded. Such
adversaries are only required to have bounded quantum storage when
Step \ref{bound} in \Randlqot\ is reached; before and after that, the
adversary can store and carry out arbitrary quantum computations
involving any number of qubits.  
Let $\mathfrak{R}_{q}$ denote the set of all possible quantum
  dishonest receivers $\dR$ in \Randlqot\ which
  have quantum memory of size at most $q$ when step~\ref{bound}
  is reached. 
We stress once more that apart from the restriction on the size
of the quantum memory available to the adversary, no other assumption
is made. In particular, the adversary is not assumed to be
computationally bounded and the size of his classical memory is not
restricted.

\begin{theorem}\label{thm:OT}




  \Randqot$^{\ell}$ is $\varepsilon$-sender-secure against
  $\mathfrak{R}_{q}$ for a negligible (in $n$) $\varepsilon$ if
  $n/4 - 2 \ell - q \in \Omega(n)$.
\end{theorem}
For improved readability, we merely give a sketch of the proof; the
formal proof that takes care of all the $\varepsilon$'s is given in
\fv{the full version~\cite{DFRSS07full}}{Appendix~\ref{app:OT}}.
\begin{proof}[sketch]
  It remains to show sender-security.  Let $X$ be the random variable
  that describes the sender's choice of $x$, where we understand the
  distribution of $X$ to be conditioned on the classical information
  that $\dR$ obtained by measuring all but $\gamma n$ qubits. A
  standard purification argument, that was also used
  in~\cite{DFSS05}, shows that the same $X$ can be obtained by
  measuring a quantum state in basis $\theta \in_R \set{+,\times}^n$,
  described by the random variable $\Theta$: for each qubit
  $\ket{x_i}_{\theta_i}$ the sender $\S$ is instructed to send to
  $\R$, $\S$ instead prepares an EPR pair \smash{$\ket{\Phi} =
    \frac{1}{\sqrt{2}}(\ket{00}+\ket{11})$} and sends one part to $\R$
  while keeping the other, and when Step~\ref{bound} is reached, $\S$
  measures her qubits.
  
  The uncertainty relation, Theorem~\ref{cor:uncertainty}, implies
  that the smooth min-entropy of $X$ given $\Theta$ is approximately
  $n/2$.  Let now $X_0$ and $X_1$ be the two substrings of $X$
  consisting of the bits encoded in the basis $+$ or $\times$,
  respectively. Then the Min-Entropy-Splitting Lemma, or, more precisely,
  Corollary~\ref{cor:ESL} implies the existence of a binary $C'$ such
  that $X_{1-C'}$ has approximately $n/4$ bits of smooth min-entropy
  given $\Theta$ and $C'$. From the random and independent choice of
  the hash functions $F_0, F_1$ and from the Chain Rule,
  Lemma~\ref{lem:chain}, it follows that $X_{1-C'}$ has still about
  $n/4 - \ell$ bits of smooth min-entropy when conditioning on
  $\Theta, C', F_{C'}$ and $F_{C'}(X_{C'})$. The Privacy Amplification
  Theorem~\ref{thm:pasmooth}, then guarantees that $S_{1-C'}
  = F_{1-C'}(X_{1-C'})$ is close to random, given $\Theta, C', F_{C'},
  S_{C'}, F_{1-C'}$ and $\dR$'s quantum state of size $q$, if
  $n/4 - 2\ell - q$ is positive and linear in $n$. \qed
\end{proof}
We note that by adapting recent and more advanced
techniques~\cite{Wullschleger07} to the quantum case, the security of
\Randqot$^{\ell}$ can be proven against $\mathfrak{R}_q$ if $n/4 - \ell - q \in \Omega(n)$.

\section{Application: Quantum Bit Commitment}\label{sec:com}

The binding criterion for classical commitments usually requires that
after the committing phase and for any dishonest committer, there
exists a unique bit $b' \in \{0,1\}$ that can only be opened with negligible
probability.  In the quantum world, this approach appears to be problematic since if 
the commitment is unconditionally concealing,
the committer can place himself in a superposition of committing to 0
and 1 and only later make a measurement that fixes the choice.  For
this reason, the previous standard approach (see e.g.~\cite{DMS00})
was to use a weaker binding condition only requiring that the
probabilities $p_0$ and $p_1$ (to successfully open $b=0$ and $b=1$
respectively), satisfy $p_0+p_1\lesssim 1$.  The bit commitment scheme
proposed in~\cite{DFSS05} was shown to be binding in this weak sense.
However, we argue that this weak notion is not really satisfactory. 
A 
shortcoming of
this notion is that committing bit by bit is not guaranteed to yield a secure
string commitment---the argument that one is tempted to use requires
independence of the $p_{b}$'s between the different executions, which
in general does not hold. 

We now argue that this notion is {\em unnecessarily} weak, at least in some cases, and in particular in the case of commitments in the bounded-quantum-storage model where the dishonest committer is forced to do some partial measurement and where we assume honest parties to produce only classical output (by measuring their entire quantum state). 
Technically, this means that for any dishonest committer $\tilde{\sf C}$, the joint state of the honest verifier and of $\tilde{\sf C}$ after the commit phase is a ccq-state \smash{$\rho_{V Z \tilde{\sf C}} = \sum_{v,z} P_{VZ}(v,z) \proj{v} \otimes \proj{z} \otimes \rho_{\tilde{\sf C}}^{v,z}$}, where the first register contains the verifier's (classical) output and the remaining two registers contain $\tilde{\sf C}$'s (partially classical) output. 
We propose the following definition. 

\begin{definition}\label{def:binding}
A commitment scheme in the bounded-quantum-storage model is called {\em $\eps$-binding}, if for every (dishonest) committer $\tilde{\sf C}$, inducing a joint state $\rho_{V Z \tilde{\sf C}}$ after the commit phase, there exists a classical binary
  random variable $B'$, given by its conditional distribution $P_{B'|VZ}$, such that for $b=0$ and $b=1$ the state \smash{$\rho_{V Z \tilde{\sf C}}^{b} = \sum_v P_{VZ|B'}(v,z|b) \proj{v} \otimes \proj{z} \otimes \rho_{\tilde{\sf C}}^{v,z}$} satisfies the following condition. When executing the opening phase on the state $\rho_{V \tilde{\sf C}}^{b}$, for any strategy of \smash{$\tilde{\sf C}$}, the honest verifier accepts an opening to $1-b$ with probability at most $\eps$. 
\end{definition}
It is easy to see that the binding property as defined here implies the above discussed weak version, namely $p_b \leq P_{B'}(b) + P_{B'}(1-b)\eps$ and thus $p_0 + p_1 \leq 1 + \eps$. 
Furthermore, it is straightforward to see that this stronger notion allows for a formal proof of the obvious reduction of a string to a bit commitment by committing bit-wise: the $i$-th
execution of the bit commitment scheme guarantees a random variable
$B'_i$, defined by $P_{B'_i|V_i Z}$, such that the committer cannot open the $i$-th bit commitment
to $1-B'_i$, and thus there exists a random variable $S'$, namely $S'
= (B'_1,\ldots,B'_m)$ defined by $P_{B'_1\cdots B'_m|V_1\cdots V_m Z} = \prod_i P_{B'_i|V_i Z}$, such that for any opening strategy, the committer cannot open the list of
commitments to any other string than $S'$.

We show in the following that the quantum bit-commitment
scheme \fv{\comm\ }{} from~\cite{DFSS05} fulfills the stronger notion of binding from
Definition~\ref{def:binding} above. For convenience, the protocol
\comm\ is reproduced in Fig.~\ref{fig:comm} below. 
Let $\mathfrak{C}_{q}$ denote the set of all possible quantum
dishonest committers \smash{$\dC$} in \comm\ which have quantum memory of size
at most $q$ at the start of the opening phase (step~\ref{it:bound}). Then the following holds.

\begin{theorem}
  The quantum bit-commitment scheme \comm\ is $\eps$-binding
  according to Definition~\ref{def:binding} against $\mathfrak{C}_{q}$
  for a negligible (in $n$) $\eps$ if $n/4 - q \in \Omega(n)$.
\end{theorem}

\begin{myfigure}{h}
\begin{myprotocol}[\small Let $b$ be the bit $\C$ want to commit to.]{\comm}
 \item $\V$ picks $x \in_R \nbit$ and $\theta \in_R \{+,\times \}^n$,
and sends $\ket{x_1}_{\theta_1},
   \ket{x_2}_{\theta_2}, \ldots, \ket{x_n}_{\theta_n}$ to $\C$.
 \item $\C$ measures all qubits in basis
   $[+,\times ]_b$ to commit to $b$. Let $x' \in \nbit$ be the result.
 \item\label{it:bound} To open the commitment, $\C$ sends $b$ and $x'$ to~$\V$.
 \item $\V$ accepts if and only if $x_i= x_i'$ for all those $i$ where $\theta_i = [+,\times ]_b$. 
\end{myprotocol}
\caption{Protocol \comm\ for commitment.}\label{fig:comm}
\end{myfigure}

\begin{proof}[Sketch]
  By considering a purified version of the scheme and using the
  uncertainty relation, one can argue that $X$ has (smooth)
  min-entropy about $n/2$ given $\Theta$.  The Min-Entropy-Splitting
  Lemma implies that there exists $B'$ such that $X_{1-B'}$ has smooth
  min-entropy about $n/4$ given $\Theta$ and $B'$. Privacy
  amplification implies that $F(X_{1-B'})$ is close to random given
  $\Theta, B', F$ and $\dC$'s quantum register of size $q$,
  where $F$ is a \univ one-bit-output hash function. This
  implies that $\dC$ cannot guess $X_{1-B'}$ except with small probability.\qed
\end{proof}

\section{Application: Quantum Key Distribution 
} \label{sec:qkd}
Let $\cB$ be a set of orthonormal bases on a Hilbert space $\cH_d$, and
assume that the basis vectors of each basis $\vartheta \in \cB$ are
parametrized by the elements of some fixed set $\cX$. We then
consider QKD protocols consisting of the steps described in Fig.~\ref{fig:QKDShape}.
Note that the quantum channel is only used in the preparation step.
Afterwards, the communication\fv{}{between Alice and Bob}
is only classical (over an authentic channel). 
\begin{myfigure}{h}
\begin{myprotocol}[\small let $N \in \bbN$ be arbitrary]{One-Way QKD}
\item \emph{Preparation:} For $i=1 \ldots N$, Alice chooses at random
  a basis $\vartheta_i \in \cB$ and a random element $X_i \in \cX$.
  She encodes $X_i$ into the state of a quantum system (e.g., a
  photon) according to the basis $\vartheta_i$ and sends this system
  to Bob.  Bob measures each of the states he receives according to a
  randomly chosen basis $\vartheta'_i$ and stores the outcome $Y_i$ of
  this measurement.
\item \emph{Sifting:} Alice and Bob publicly announce their choices of
  bases and keep their data at position $i$ only if $\vartheta_i =
  \vartheta'_i$. In the following, we denote by $X$ and $Y$ the
  concatenation of the remaining data $X_i$ and $Y_i$, respectively.
  $X$ and $Y$ are sometimes called the \emph{sifted raw key}.
\item \emph{Error correction:} Alice computes some error correction
  information $C$ depending on $X$ and sends $C$ to Bob.  Bob computes
  a guess $\hat{X}$ for Alice's string $X$, using $C$ and $Y$.
\item \emph{Privacy amplification:} Alice chooses at random a function
  $f$ from a two-universal family of hash functions and announces $f$
  to Bob. Alice and Bob then compute the final key by applying $f$ to
  their respective strings $X$ and $\hat{X}$. 
\end{myprotocol}
\caption{General form for {\em one-way} QKD protocols.}\label{fig:QKDShape}
\end{myfigure}

As shown in~\cite{Renner05} (Lemma~6.4.1), the length $\ell$ of the
secret key that can be generated in the privacy amplification step of
the protocol described above is given by\footnote{The approximation in
  this and the following equations holds up to some small additive
  value which depends logarithmically on the desired security $\ep$ of
  the final key.}
\[
\ell \approx \hiee{X \mid \rs} - \hmax(C) \enspace ,
\]
where $\rs$ denotes the (quantum) system containing all the
information Eve might have gained during the preparation step of the
protocol and where $\hmax(C)$ is the number of error correction bits
sent from Alice to Bob. Note that this formula can be seen as a
generalization of the well known expression by Csisz\'{a}r and
K\"{o}rner for classical key agreement~\cite{CK78}.

Let us now assume that Eve's system $\rs$ can be decomposed into a
classical part $Z$ and a purely quantum part $\rs'$. Then, using the
chain rule (Lemma~3.2.9 in~\cite{Renner05}), we find
\[
  \ell 
\approx 
  \hiee{X \mid Z \rs'} - \hmax(C) 
\gtrsim
  \hiee{X \mid Z} - \hmax(\rs') - \hmax(C) \enspace .
\]
Because, during the preparation step, Eve does not know the encoding
bases which are chosen at random from the set $\cB$, we can apply our
uncertainty relation (Theorem~\ref{thm:genrel}) to get a lower bound for
the min-entropy of $X$ conditioned on Eve's classical information $Z$,
i.e.,
$\hiee{X \mid Z} \geq M h$,
where $M$ denotes the length of the sifted raw key $X$ and $h$ is the
average entropic uncertainty bound for $\cB$. Let $q$ be the bound on
the size of Eve's quantum memory $\rs'$. Moreover, let $e$ be the
average amount of error correction information that Alice has to send
to Bob per symbol of the sifted raw key $X$. Then
$
  \ell 
\gtrsim 
  M (h-e) - q \enspace .
$
Hence, if the memory bound only grows sublinearly in the length $M$ of
the sifted raw key, then the \emph{key rate}, i.e., the number of key
bits generated per bit of the sifted raw key, is lower bounded by
$$
  \mathrm{rate}
\geq
  h-e \enspace .
$$

\paragraph{The Binary-Channel Setting.}
For a binary channel (where $\cH$ has dimension two), the average
amount of error correction information $e$ is given by the binary
Shannon entropy\footnote{This value of $e$ is only achieved if an
  optimal error-correction scheme is used. In practical
  implementations, the value of $e$ might be slightly larger.} $\Hb(p)
= -\big(p\log(p)+(1-p)\log(1-p)\big)$, where $p$ is the bit-flip
probability of the quantum channel (for classical bits encoded
according to some orthonormal basis as described above). The
achievable key rate of a QKD protocol using a binary quantum channel
is thus given by $ \mathrm{rate}_{\mathrm{binary}} \geq h - \Hb(p) $.
Summing up, we have derived the following theorem.


\begin{theorem}
  Let $\cB$ be a set of orthonormal bases of $\cH_2$ with average
  entropic uncertainty bound $h$. Then, a \emph{one-way} QKD-protocol
  as in Fig.~\ref{fig:QKDShape} produces a secure key against
  eavesdroppers whose quantum-memory size is sublinear in the length
  of the raw key (i.e., sublinear in the number of qubits sent from
  Alice to Bob) at a positive rate as long as the bit-flip probability
  $p$ fulfills
  $\Hb(p) < h$.
\end{theorem}

For the BB84 protocol, we have $h = \frac{1}{2}$ and $\Hb(p) < \frac{1}{2}$
is satisfied as long as $p \leq 11\%$.  This bound coincides with the
known bound for security against an unbounded adversary. So, the
memory-bound does not give an advantage here.\footnote{Note, however,
  that the analysis given here might not be optimal.}

The situation is different for the six-state protocol where $h =
\frac{2}{3}$. 
In this case, security against memory-bounded adversaries is
guaranteed (i.e. $\Hb(p) < \frac{2}{3}$) as long as $p \leq 17\%$. If
one requires security against an unbounded adversary, the threshold for
the same protocol lies below $13\%$, and even the best known QKD
protocol on binary channels with one-way classical post-processing can
only tolerate noise up to roughly $14.1\%$~\cite{RGK05}. It has also
been shown that, in the unbounded model, no such protocol can tolerate
an error rate of more than~$16.3\%$.

The performance of QKD protocols against quantum-memory bounded
eavesdroppers can be improved further by making the choice of the
encoding bases more random. For example, they might be chosen from the
set of all possible orthonormal bases on a two-dimensional Hilbert
space.  As shown in \fv{the full
  version~\cite{DFRSS07full}}{Appendix~\ref{app:uncertbound}}, the
average entropic uncertainty bound is then given by $h \approx 0.72$
and 
$\Hb(p) < 0.72$ is satisfied if $p \lesssim 20\%$.  For an
unbounded adversary, the thresholds are the same as for the six-state
protocol (i.e., $14.1\%$ for the best known one-way protocol).

\section{Open Problems}
It is interesting to investigate whether the uncertainty relation
(Theorem~\ref{thm:genrel}) still holds if the measurement bases
$(\Theta_1,\ldots,\Theta_n)$ are randomly chosen from a relatively
small subset of $\cB^n$ (rather than from the entire set $\cB^n$).
Such an extension would reduce the amount of randomness that is needed
in applications. In particular, in the context of QKD with
quantum-memory-bounded eavesdroppers, it would allow for more
efficient protocols that use a relatively short initial secret key in
order to select the bases for the preparation and measurement of the
states and, hence, avoid the sifting step.

Another open problem is to consider protocols using higher-dimensional
quantum systems. The results described in
Appendix~\ref{app:uncertbound} show that for $d$-dimensional systems,
the average entropic uncertainty bound converges to $\log{d}$ for
large $d$. The maximal tolerated channel noise might thus be higher for
such protocols (depending on the noise model for higher-dimensional
quantum channels).





\bibliographystyle{abbrv}

\bibliography{crypto,qip,procs}

\RemoveAppendix{
\begin{appendix}

\section{Proofs}

\subsection{Proof of Theorem~\ref{thm:hmin} (Uncertainty Relation)} \label{app:proofhmin}
Define $\prei[i]{Z} \assign (Z_1,\ldots,Z_i)$ for any $i \in
\set{1,\ldots,n}$, and similarly for other sequences.  We want to show
that \smash{$\Pr\big[P_{\prei[n]{Z}}(\prei[n]{Z}) \geq
  2^{-(h-2\lambda)n}\big] \leq \varepsilon$} for $\varepsilon$ as
claimed in Theorem~\ref{thm:hmin}.  This means that
$P_{\prei[n]{Z}}(\prei[n]{z})$ is smaller than $2^{-(h-2\lambda)n}$
except with probability at most $\varepsilon$ (over the choice of
$\prei[n]{z}$), and therefore implies the claim
\mbox{$H_{\infty}^{\varepsilon}(Z^n) \geq (h-2\lambda)n$}
by the definition of smooth min-entropy.
Note that
$P_{\prei[n]{Z}}(\prei[n]{Z}) \geq 2^{-(h-2\lambda)n}$ is equivalent to
\begin{equation}\label{eq:bound1}
  \sum_{i=1}^n \Big( \log\big( P_{Z_i \mid \prei{Z}}(Z_i \mid \prei{Z}) \big) + h \Big) \geq 2 \lambda n
\end{equation}
which is of suitable form to apply Azuma's inequality (Lemma~\ref{lem:azuma}). 

Consider first an arbitrary sequence $S_1,\ldots,S_n$ of real-valued
random variables. We assume the $S_i$'s to be either all positive or
all negative. Define a new sequence $R_1,\ldots,R_n$ of random
variables by putting $R_i := S_i - \E[S_i \mid \prei{S}]$. It is
straightforward to verify that $\E[R_i \mid \prei{R}] = 0$, i.e.,
$R_1,\ldots,R_n$ forms a martingale difference sequence. Thus, if
$|S_i| \leq c$ for some $c$ (and any $i$), and thus $|R_i| \leq c$,
Azuma's inequality guarantees that
\begin{equation}\label{eq:bound2}
\Pr\left[\sum_{i=1}^n \Big(S_i - \E\big[S_i \mid \prei{S}\big]\Big) \geq \lambda n\right] \leq \exp\left(-\frac{\lambda^2 n}{2 c^2}\right) \, .
\end{equation}
We now put $S_i := \log P_{Z_i \mid \prei{Z}}(Z_i \mid \prei{Z})$ for
$i=1,\ldots,n$. Note that $S_1,\ldots,S_n \leq 0$. It is easy to see that
the bound on the conditional entropy of $Z_i$ from
Theorem~\ref{thm:hmin} implies that $\E[S_i \mid \prei{S}] \leq -h$.  Indeed,
for any $\prei{z} \in {\cal Z}^{i-1}$, we have $\E\big[\log
P_{Z_i \mid \prei{Z}}(Z_i \mid \prei{Z}) \mid \prei{Z}\!=\!\prei{z}\big] = -
\H(Z_i \mid \prei{Z}\!=\!\prei{z}) \leq - h$, and thus for any subset $\cal E$
of ${\cal Z}^{i-1}$, and in particular for the set of $\prei{z}$'s
which map to a given $\prei{s}$, it holds that
\begin{align} \nonumber
\E\big[S_i \mid \prei{Z}\!\in\!{\cal E}\big] &=
\sum_{\prei{z} \in \cal
  E}\!\! P_{\prei{Z} \mid \prei{Z}\in{\cal E}}(\prei{z}) \cdot
  \E\big[\log P_{Z_i \mid \prei{Z}}(Z_i \mid \prei{Z}) \mid
  \prei{Z}\!=\!\prei{z}\big]\\
 &\leq - h \, . \label{eq:bound3}
\end{align}
As a consequence, the bound on the probability of (\ref{eq:bound2}) in
particular bounds the probability of the event (\ref{eq:bound1}), even
with $\lambda n$ instead of $2 \lambda n$. A problem though is that we
have no upper bound $c$ on the $|S_i|$'s.  Because of that we now
consider a modified sequence $\tilde{S}_1,\ldots,\tilde{S}_n$ defined
by $\tilde{S}_i := \log P_{Z_i \mid \prei{Z}}(Z_i \mid \prei{Z})$ if
$P_{Z_i \mid \prei{Z}}(Z_i \mid \prei{Z}) \geq \delta$ and
$\tilde{S}_i := 0$ otherwise, where $\delta > 0$ will be determined
later.  This gives us a bound like (\ref{eq:bound2}) but with an
explicit $c$, namely $c = \log(1/\delta)$. Below, we will argue that
$\E\big[\tilde{S}_i \mid \prei{\tilde{S}}\big]-\E\big[S_i \mid
\prei{\tilde{S}}\big] \leq \lambda$ by the right choice of $\delta$;
the claim then follows from observing that
\begin{align*}
\tilde{S}_i - \E\big[\tilde{S}_i \mid \prei{\tilde{S}}\big] &\geq S_i -
\E\big[\tilde{S}_i \mid \prei{\tilde{S}}\big]\\ 
&\geq S_i - \E\big[S_i \mid \prei{\tilde{S}}\big] - \lambda\\
&\geq S_i + h - \lambda,
\end{align*}
where the last inequality follows from (\ref{eq:bound3}).  Regarding
the claim $\E\big[\tilde{S}_i \mid \prei{\tilde{S}}\big]-\E\big[S_i
\mid \prei{\tilde{S}}\big] \leq \lambda$, using a similar argument as
for (\ref{eq:bound3}), it suffices to show that $\E\big[\tilde{S}_i
\mid \prei{\tilde{Z}}\!=\!\prei{z}\big]-\E\big[S_i \mid
\prei{\tilde{Z}}\!=\!\prei{z}\big] \leq \lambda$ for any $\prei{z}$:
\begin{align*}
\E\big[\tilde{S}_i \mid \prei{\tilde{Z}}\!=\!\prei{z}\big]-
\E\big[S_i \mid \prei{\tilde{Z}}\!=\!\prei{z}\big]
  &= - \sum_{z_i} P_{Z_i |\prei{Z}}(z_i \mid \prei{z})
  \log(P_{Z_i | \prei{Z}}(z_i \mid \prei{z}))\\
  &\leq |\mathcal{Z}| \delta \log ( 1/ \delta)
\end{align*}
where the summation is over all $z_i \in {\cal Z}$ with $P_{Z_i | \prei{Z}}(z_i \mid \prei{z}) <
    \delta$, and 
where the inequality holds as long as $\delta \leq 1/e$, as can
easily be verified. Thus, we let $0<\delta<1/e$ be such that
$|\mathcal{Z}| \delta \log(1/\delta) = \lambda$.  Using
Lemma~\ref{lem:delta} below, we have that $\delta > \frac{\lambda /
  |\mathcal{Z}|}{4 \log{(|\mathcal{Z}| / \lambda})}$ and derive
that $c^2 = \log(1/\delta)^2 = \lambda^2/(\delta|{\cal Z}|)^2 < 16
\log(|\mathcal{Z}|/\lambda)^2$, which gives us the claimed bound
$\varepsilon$ on the probability.
\qed

\begin{lemma} \label{lem:delta}
  For any $0 < x < 1/e$ such that $y \assign x \log(1/x) < 1/4$,
  it holds that $x > \frac{y}{4 \log(1/y)}$.
\end{lemma}
\begin{proof}
  Define the function $x \mapsto f(x) = x \log(1/x)$. It holds that
  $f'(x) = \frac{d}{dx}f(x) = \log(1/x)-\log e$, which shows that $f$
  is bijective in the interval $(0,1/e)$, and thus the inverse
  function $f^{-1}(y)$ is well defined for $y \in (0,\log(e)/e)$,
  which contains the interval $(0,1/4)$. We are going to show that
  $f^{-1}(y) > g(y)$ for all $y \in (0,1/4)$, where $g(y) = \frac{y}{4
    \log(1/y)}$. Since both $f^{-1}(y)$ and $g(y)$ converge to 0 for
  $y \rightarrow 0$, it suffices to show that $\frac{d}{dy} f^{-1}(y)
  > \frac{d}{dy} g(y)$; respectively, we will compare their
  reciprocals. For any $x \in (0,1/e)$ such that $y = f(x) = x
  \log(1/x) < 1/4$
$$
\frac{1}{\frac{d}{dy} f^{-1}(y)} = f'(f^{-1}(y)) = \log(1/x)-\log(e)
$$
and
$$
\frac{d}{dy} g(y) = \frac{1}{4} \bigg( \frac{1}{\log(1/y)} + \frac{1}{\ln(2) \log(1/y)^2} \bigg)
$$
such that 
\begin{align*}
\frac{1}{\frac{d}{dy} g(y)} = 4 \, \frac{\ln(2)\log(1/y)^2}{\ln(2) \log(1/y) + 1} 
= 4 \, \frac{\log(1/y)}{1+\frac{1}{\ln(2) \log(1/y)}}
& > 2 \log\Big(\frac{1}{y}\Big) 
= 2 \log\Big(\frac{1}{x \log(1/x)}\Big) \\[0.7ex]
&= 2\big(\log(1/x) - \log\log(1/x)\big)
\end{align*} 
where for the inequality we are using that $y < 1/4$ so that $\ln(2) \log(1/y) > 2\ln(2) = \ln(4) > 1$. 
Defining the function
$$
h(z) \assign z - 2\log(z) + \log(e)
$$
and showing that $h(z) > 0$ for all $z>0$ finishes the proof, as then
$$
0 < h\big(\log(1/x)\big) \leq  \frac{1}{\frac{d}{dy} g(y)} - \frac{1}{\frac{d}{dy} f^{-1}(y)}
$$
which was to be shown. 
For this last claim, note that $h(z) \rightarrow \infty$ for $z
\rightarrow 0$ and for $z \rightarrow \infty$, and thus the global
minimum is at $z_0$ with $h'(z_0) = 0$. $h'(z) = 1 - 2/(\ln(2)z)$ and
thus $z_0 = 2/\ln(2) = 2\log(e)$, and hence the minimum of $h(z)$ equals
$h(z_0) = 3 \log(e) - 2\log\big(2\log(e)\big)$, which turns out to be positive. 
\end{proof}

\subsection{Proof of Theorem~\ref{thm:pasmooth} 
(Privacy Amplification With Classical Conditioning)
}\label{app:derivePA}
\newcommand{\hminee}{\H_{\rm min}^{\varepsilon}}

In this section, we adopt the slightly more advanced notation
from~\cite{Renner05} in order to derive Theorem~\ref{thm:pasmooth}
from Corollary~5.6.1 in~\cite{Renner05}.  In our case, the quantum
register $B$ from Corollary~5.6.1 consists of a classical part $U$ and
a quantum part $\rs$. Denoting by $\sigma_Q$ the fully mixed state on
the image of $\rho_Q$, we only need to consider the term in the
exponent to derive Theorem~\ref{thm:pasmooth} as follows
\begin{align}
 \hminee(\rho_{XUQ} \mid UQ) \nonumber &\geq \hminee(\rho_{XUQ} \mid
  \rho_{U} \otimes \sigma_Q) \nonumber\\
&\geq \hminee(\rho_{XUQ} \mid \rho_{U}) - \H_{\rm max}(\rho_{Q}) \label{eq:chainrule}\\
&\geq \hminee(\rho_{XU} \mid \rho_U) - \H_{\rm max}(\rho_{Q}) \label{eq:lemma}\\
&= \hiee{X \mid U} - q. \nonumber
\end{align}
The first inequality follows by Definition~3.1.2 in~\cite{Renner05} of
$\hminee$ as supremum over all $\sigma_{UQ}$. Inequality
\eqref{eq:chainrule} is the chain rule for smooth min-entropy
(Lemma~3.2.9 in~\cite{Renner05}). Inequality~\eqref{eq:lemma} uses
that the smooth min-entropy cannot decrease when dropping the quantum
register which is proven in Lemma~\ref{lem:dropquantum} below. The
last step follows by observing that the quantum quantities defined
in~\cite{Renner05} correspond to the notions used in this paper
accordingly (see Remark~3.1.4 in~\cite{Renner05}). \qed

\begin{lemma} \label{lem:dropquantum_notsmooth}
Let $\rho_{XUQ} \in \dens{{\cal H}_X \otimes {\cal H}_U \otimes {\cal
    H}_Q}$ be classical on ${\cal H}_X \otimes {\cal H}_U$. Then
$$\H_{\rm min}(\rho_{XUQ} \mid \rho_U) \geq \H_{\rm min}(\rho_{XU} \mid \rho_U).$$
\end{lemma}
\begin{proof}
  For $\lambda \assign 2^{-\H_{\rm min}(\rho_{XU} \mid \rho_U)}$, we
  have by Definition~3.1.1 in~\cite{Renner05} that $\lambda \cdot
  \id_X \otimes \rho_U - \rho_{XU} \geq 0$. Using that both $X$ and
  $U$ are classical, we derive that for all $x,u$, it holds $\lambda
  \cdot p_u - p_{xu} \geq 0$, where $p_u$ and $p_{xu}$ are shortcuts
  for the probabilities $P_U(u)$ and $P_{XU}(x,u)$.  Let the
  normalized conditional operator $\ol{\rho}_Q^{x,u}$ be defined as in
  Sect.~2.1.3 of~\cite{Renner05}. Then,
\[\sum_{x,u} \lambda \cdot p_u \ol{\rho}_Q^{x,u} \otimes \proj{xu} - p_{xu}
\ol{\rho}_Q^{x,u} \otimes \proj{xu} \geq 0.\]
Because of $\ol{\rho}_Q^{x,u} \leq \id_Q$, we get
\[\sum_{x,u} \lambda \cdot p_u \id_Q  \otimes \proj{xu} - p_{xu}
\ol{\rho}_Q^{x,u} \otimes \proj{xu} \geq 0.\]
Therefore, it holds $\lambda \cdot \id_{QX} \otimes \rho_U - \rho_{QXU} \geq 0$,
from which follows by definition that $\H_{\rm min}(\rho_{XUQ} \mid \rho_U) \geq
-\log(\lambda)$.
\end{proof}

\begin{lemma} \label{lem:dropquantum}
Let $\rho_{XUQ} \in \dens{{\cal H}_X \otimes {\cal H}_U \otimes {\cal
    H}_Q}$ be classical on ${\cal H}_X \otimes {\cal H}_U$ and let
    $\varepsilon \geq 0$. Then
$$\hminee(\rho_{XUQ} \mid \rho_U) \geq \hminee(\rho_{XU} \mid \rho_U).$$
\end{lemma}
\begin{proof}
  After Remark~3.2.4 in~\cite{Renner05}, there exists $\sigma_{XU} \in
  \ball{\varepsilon}(\rho_{XU})$ classical on ${\cal H}_X \otimes
  {\cal H}_U$ such that $\hminee(\rho_{XU} \mid \rho_U) = \H_{\rm
    min}(\sigma_{XU} \mid \sigma_U)$. Because both $X$ and $U$ are
  classical, we can write $\sigma_{XU} = \sum_{x,u} p_{xu} \proj{xu}$
  and extend it to obtain $\sigma_{XUQ} \assign \sum_{x,u} p_{xu}
  \proj{xu} \otimes \ol{\rho}_Q^{x,u}$.
  Lemma~\ref{lem:dropquantum_notsmooth} above yields $\H_{\rm
    min}(\sigma_{XU} \mid \sigma_U) \leq \H_{\rm min}(\sigma_{XUQ}
  \mid \sigma_U)$.  We have by construction that $\delta(\sigma_{XUQ},
  \rho_{XUQ}) = \delta(\sigma_{XU},\rho_{XU}) \leq \varepsilon$.
  Therefore, $\sigma_{XUQ} \in \ball{\varepsilon}(\rho_{XUQ})$ and
  $\H_{\rm min}(\sigma_{XUQ} \mid \sigma_U) \leq \hminee(\rho_{XUQ}
  \mid \rho_U).$
\end{proof}

\subsection{Proof of Lemma~\ref{lemma:ESL} (Min-Entropy-Splitting Lemma) }\label{app:ESL}

In the following, we give the proof for $\varepsilon = 0$, i.e., for ordinary
(non-smooth) min-entropy. The general claim for smooth min-entropy
follows immediately by observing that the same argument also works for
non-normalized distributions with a total probability smaller than 1.

We extend the probability distribution $P_{X_0 X_1}$ as follows to
  $P_{X_0 X_1 C}$. Let $C=1$ if $P_{X_1}(X_1) \geq 2^{-\alpha/2}$ and $C=0$ otherwise.
  We have that for all $x_1$, $P_{X_1 C}(x_1,0)$ either vanishes or is equal
  to $P_{X_1}(x_1)$. In any case, $P_{X_1 C}(x_1,0) < 2^{-\alpha/2}$.
  
  On the other hand, for all $x_1$ with $P_{X_1 C}(x_1,1)>0$, we have
  that $P_{X_1 C}(x_1,1)=P_{X_1}(x_1) \geq 2^{-\alpha/2}$ and
  therefore, for all $x_0$,
\begin{equation*} 
 P_{X_0 X_1 C}(x_0,x_1,1) \leq 2^{-\alpha} =2^{-\alpha/2} \cdot
 2^{-\alpha/2} \leq 2^{-\alpha/2} P_{X_1}(x_1).
\end{equation*}
Summing over all $x_1$ with $P_{X_0 X_1 C}(x_0,x_1,1) > 0$, and thus with $P_{X_1 C}(x_1,1) > 0$, results in
$$
P_{X_0 C}(x_0,1) \leq \sum_{x_1} 2^{-\alpha/2} P_{X_1}(x_1)
\leq 2^{-\alpha/2}.
$$
This shows that $P_{X_{1-C} C}(x,c) \leq 2^{-\alpha/2}$ for all $x,c$.
\subsection{Proof of Theorem~\ref{thm:OT} (Sender-Security of the OT Scheme)} \label{app:OT}

First, we consider a purified version of \Randlqot, \eprRandlqot\ in
Fig.~\ref{fig:eprRandlqot}, where for each qubit
$\ket{x_i}_{\theta_i}$ the sender $\S$ is instructed to send to the
receiver, $\S$ instead prepares an EPR pair \smash{$\ket{\Phi} =
  \frac{1}{\sqrt{2}}(\ket{00}+\ket{11})$}, and sends one part to the
receiver while keeping the other. Only when Step~\ref{bound} is
reached and $\dR$'s quantum memory is bound to $\gamma n$ qubits, $\S$
measures her qubits in basis $\theta \in_R \set{+,\times}^n$. It is
easy to see that for any $\dR$, \eprRandlqot\ is equivalent to the
original \Randlqot, and it suffices to prove sender-security for the
former.  Indeed, $\S$'s choices of $\theta$ and $\hf_0,\hf_1$,
together with the measurements all commute with $\R$'s actions.
Therefore, they can be performed right after Step 1 with no change for
$\R$'s view. Modifying \eprRandlqot\ that way results in \Randlqot.  A
similar approach was used in~\cite{DFSS05}, or in~\cite{SP00} in the
context of the BB84 quantum key distribution scheme.

\begin{myfigure}{h}
\begin{myprotocol}{\eprRandlqot}
\item $\S$ prepares $n$ EPR pairs each in state 
      $\ket{\Omega}=\frac{1}{\sqrt{2}}(\ket{00}+\ket{11})$, 
and sends one half of each pair to $\R$ and keeps the other
  halves.\label{rec}
\item $\R$ measures all qubits in basis $[+,\times]_{c}$. Let $x' \in
  \{0,1\}^n$ be the result.
\item $\S$ picks random $\theta \in_R \{+,\times \}^n$, and she
  measures the $i$th qubit in basis $\theta_i$. Let $x\in\{0,1\}^n$ be
  the outcome. $\S$ picks two hash functions $\hf_0, \hf_1 \in_R \UH$,
  announces $\theta$ and $\hf_0, \hf_1$ to $\R$, and outputs $s_0
  \assign \hf_0(x |_{I_0})$ and $s_1 \assign \hf_1(x |_{I_1})$ where
  $I_b \assign \Set{i}{\theta_i \!=\! [+,\times]_b}$.
\item $\R$ outputs $s_{c} = \hf_{c}(x' |_{I_{c}})$.
\end{myprotocol}
\caption{Protocol for EPR-based \boldmath\RandlStringOT.}\label{fig:eprRandlqot}
\end{myfigure}

Consider the common quantum state in \eprRandlqot\ after $\dR$ has
measured all but $\gamma n$ of his qubits. Let $X$ be the random
variable that describes the outcome of the sender measuring her part
of the state in random basis $\Theta$, and let $\rs$ be the random
state that describes $\dR$'s part of the state.  Also, let $\Hf_0$ and
$\Hf_1$ be the random variables that describe the random and
independent choices of $\hf_0,\hf_1 \in \UH$.  Finally, let $X_b$ be
$X_b = X|_{\Set{i}{\Theta_i = [+,\times]_b}}$ (padded with zeros so it
makes sense to apply $\Hf_b$).

Choose $\lambda, \lambda',\kappa$ all positive, but small enough such
  that $\gamma n \leq (1/4 - \lambda - 2 \lambda' - \kappa) n -
  2\ell - 1$. 
  From the uncertainty relation (Corollary~\ref{cor:uncertainty}), we
  know that $\hie{\ep}{X_0 X_1 \mid \Theta} \geq (1/2 - 2\lambda)n$ for $\ep$ exponentially small in $n$.
  Therefore, by Corollary~\ref{cor:ESL}, there exists a binary random
  variable $C'$ such that for $\ep'=2^{-\lambda' n}$, it holds that
\[\hie{\ep+\ep'}{X_{1-C'} \mid \Theta,C'} \geq (1/4 - \lambda - \lambda')n -1 \, .\]
We denote by the random variables
  $F_0,F_1$ the sender's choices of hash functions.  It is clear
  that we can condition on the independent $F_{C'}$ and use the chain rule
  (Lemma~\ref{lem:chain}) to obtain
\begin{align*}
  \hie{\ep+2\ep'}{X_{1-C'} &\mid \Theta F_{C'}(X_{C'}) F_{C'}, C'}\\
  & \geq \hie{\ep+2\ep'}{X_{1-C'} F_{C'}(X_{C'}) \mid \Theta F_{C'} C'} -
  \hmax(F_{C'}(X_{C'}) \mid F_{C'} C') - \lambda'n \\
&\geq (1/4 -\lambda - 2 \lambda')n - \ell -1\\
&\geq \gamma n +\ell +\kappa n,
\end{align*}
by the choice of $\lambda, \lambda', \kappa$.
 We can now apply privacy amplification in form of Theorem~\ref{thm:pasmooth}
 to obtain
\begin{align*}
 \dist(F_{1-C'}(X_{1-C'}) \mid &F_{1-C'}, \Theta F_{C'}(X_{C'}) F_{C'}C',
 \rs)\\
&\leq \frac12 2^{-\frac12 \left( \hie{\ep+2\ep'}{ X_{1-C'} \mid
 \Theta F_{C'}(X_{C'})F_{C'}C'} - \gamma n - \ell \right)} + 2
 (\ep+2\ep')\\
&\leq \frac12 2^{-\frac12 \kappa n} + 2\ep + 4\ep',
\end{align*} 
 which is negligible.\qed
 

\section{Computing the Overall Average Entropic Uncertainty Bound}\label{app:uncertbound}

Let $\cU(d)$ be the set of unitaries on $\cH_d$.  Moreover, let $d U$
be the normalized Haar measure on $\cU(d)$, i.e.,
\[
  \int_{\cU(d)} f(V U) d U 
= 
  \int_{\cU(d)} f(U V) d U 
= 
  \int_{\cU(d)} f(U) d U \ ,
\]
for any $V \in \cU(d)$ and any integrable function $f$, and $\int_{\cU(d)} dU = 1$. (Note that the
normalized Haar measure $d U$ exists and is unique.)

\def\all{\text{\rm all}}

Let $\{\omega_1, \ldots, \omega_d\}$ be a fixed orthonormal basis of
$\cH_d$, and let $\cB_{\all} = \{\vartheta_U\}_{U \in \cU(d)}$ be the
family of bases $\vartheta_U = \{U \omega_1, \ldots, U \omega_d\}$
with $U \in \cU(d)$. The set $\cB_{\all}$ consist of {\em all}
orthonormal basis of $\cH_d$. We generalize Definition~\ref{def:aeub},
the average entropic uncertainty bound for a finite set of bases, to
the {\em infinite} set $\cB_{\all}$.
\begin{definition}
We call $h_d$ an {\em overall average entropic uncertainty bound} in $\cH_d$ if every state in $\cH_d$ satisfies
  \[
     \int_{\cU(d)} \H(P_{\vartheta_U}) d U \geq h_d \ ,
  \]
  where $P_{\vartheta_U}$ is the distribution obtained by measuring the state in basis $\vartheta_U \in \cB_{\all} $. 
  \end{definition}
  
\begin{proposition} \label{prop:comph} 
For any positive integer $d$, 
  \[
  h_d = \left( \sum_{i=2}^d \frac{1}{i} \right) / \ln(2)
  \]
  is the overall average entropic uncertainty bound in $\cH_d$. It is
  attained for any pure state in $\cH_d$.
\end{proposition}
The proposition follows immediately from Formula~(14) in~\cite{JRW94}
for a pure state, i.e. $(\lambda_1,\ldots,\lambda_n)=(1,0,\ldots,0)$.
The result was originally shown in~\cite{Sykora74,Jones91}, another
proof can be found in the appendix of~\cite{JRW94}.

The following table gives some numerical values of $h_d$ for small
values of $d$.
\begin{center}
\begin{tabular}{c|cccc}
  $d$                    & $2$    & $4$    & $8$    & $16$ \\
  \hline
  $h_d$                  & $0.72$ & $1.56$ & $2.48$ & $3.43$ \\
  $\frac{h_d}{\log_2(d)}$ & $0.72$ & $0.78$ & $0.83$ & $0.86$ 
\end{tabular}
\end{center}

It is well-known that the harmonic series in
Proposition~\ref{prop:comph} diverges in the same way as $\log_2(d)$
and therefore, $\frac{h_d}{\log_2(d)}$ goes to 1 for large dimensions
$d$.

\end{appendix}
}

\end{document}